\documentclass[twocolumn,prl,amstex]{revtex4}

\usepackage{graphicx}

\def\be{\begin{equation}}
\def\ee{\end{equation}}
\def\bea{\begin{eqnarray}}
\def\eea{\end{eqnarray}}
\def\bma{\begin{mathletters}}
\def\ema{\end{mathletters}}

\def\0{\overline{0}}

\def\q0{\underline{0}}

\def\one{\leavevmode\hbox{\small1\normalsize\kern-.33em1}}

\begin{document}

%\draft \wideabs{

\title{
Quantum measurement of the degree of polarization of a light beam}

\author{M. Legr\'e$^*$, M. Wegm\"uller, N. Gisin}

\address{
GAP-Optique, University of Geneva, 20, Rue de l'\'Ecole de
M\'edecine, CH-1211 Geneva 4, Switzerland }

\date{5 May 2003}

%%%%%%%%%%%% Abstract %%%%%%%%%%%%%%%%%%%%%%%%%%%

\begin{abstract}

We demonstrate a coherent quantum measurement for the
determination of the degree of polarization (DOP). This method
allows to measure the DOP in the presence of fast polarization
state fluctuations, difficult to achieve with the typically used
polarimetric technique. A good precision of the DOP measurements
is obtained using 8 type II nonlinear crystals assembled for
spatial walk-off compensation.

\end{abstract}

%\pacs{PACS Nos. 03.67.-a, 03.65.Ud}

\maketitle

\section{I Introduction}
\label{intr}
The history of the concept of polarization of light
is fascinating and very instructive of the way science progresses,
see e.g. \cite{polar}. Today, there is a renewed interest because
of the fast developments in optics, both on the applied side for
optical communication and on the more academic side for quantum
optics. In this letter we concentrate on the {\it Degree Of
Polarization} (DOP) which is often desired to reach its maximum
value of 1, as well for close-to-ideal classical as for quantum
communication \cite{infquant}. We analyse this problem from a
quantum perspective, and then apply the gained insight to an
experimental measurement of the DOP using classical nonlinear
optics.

It is well known that depolarization is due to decoherence. A
light beam can be (partially) depolarized ($DOP<1$) for any
combination of 3 basic causes: mixture of spatial modes with
different polarization, mixture of temporal modes with different
polarization, and mixture of spectral modes with different
polarization.

Clearly, light propagating in a single-mode fiber can not suffer
from depolarization due to the first cause. Moreover, one is often
not interested in depolarization due to time-fluctuations (see
e.g. the discussion below about polarization mode dispersion).
Consequently, one would like a measurement technique providing
information on the "{\it instantaneous}" DOP of a single-mode
light beam. Note that "instantaneous" does not refer to an
infinitesimal time interval -for which polarization is not even
defined-, but to the coherence time of the signal. Measuring the
"instantaneous" DOP is a non trivial task, since classical
polarimeters measure the 4 Stokes parameters and then compute the
DOP. In other words, the usual measurement technique is an
indirect one, necessarily requiring some time to average the
intensities on the 4 detectors providing the Stokes parameters.
Let us look at this problem from a fundamental point of view,
considering the quantum nature of light. If one has only a single
photon at disposal and measures its polarization along any (linear
or elliptical) direction, one obtains one out of two possible
results. It is easy to convince oneself (and this can be made
rigorous \cite{VLPT}) that this single result provides absolutely
no information  on the DOP (not even probabilistic information,
i.e. it doesn't help at all to guess the correct DOP) of the beam
from which this photon was extracted. It is only by accumulating
several results on photons from the same beam that one can gain
some information. But accumulating results necessarily takes some
time, hence possibly the DOP measurement gets spoiled by
time-fluctuations of the state of polarization. Note that
classical linear optics does nothing else than accumulating
measurement results on individual photons, thus measuring the DOP
in an indirect way. Consequently, the only possibility to improve
DOP measurements consists in processing the photons in pairs (or
triplets, etc), i.e. accessing directly the DOP.

From quantum information theory we learned in the recent years
that {\it coherent measurements}, that is measurements represented
by self-adjoint operators whose eigenstates are entangled, do
indeed generally provide more information than successive
individual measurements \cite{MP}. This came as a surprise, since
it applies also to the case where the measured systems are not
entangled, as for the case under investigation: the photons of a
classical light beam are not entangled, but coherent measurements
do provide more information. For DOP measurement \cite{DOPJMO1},
the optimal coherent quantum measurement is represented by the
operator {\it projection on the singlet state}: \be
P_{singlet}=\frac{1}{2}(|H,V\rangle-|V,H\rangle)(\langle
H,V|-\langle V,H|) \ee

This can be understood intuitively. If light is perfectly
polarized, DOP=1, then all photons are in the same polarization
state. Consequently, the projection of any pair of photons on the
singlet state is zero (recall that the singlet state is
rotationally invariant). But if the DOP is less than unity, then
there is a finite probability that a pair of photon projects
during a measurement process onto the singlet state. Let us make
this quantitative. Let $\{S_j\}_{j=0,1,2,3}$ denote the Stokes
parameters. The polarization vector $\vec M$ on the Poincar\'e
sphere is then $M_j=\frac{S_j}{S_0}$, j=1, 2, 3, and the quantum
state of polarization is represented by the density matrix
$\rho=\frac{\one+\vec M\vec\sigma}{2}$, where $\vec\sigma$ are the
Pauli matrices. The DOP is related to the Poincar\'e vector by
DOP=$|\vec M|$. Accordingly, the probability that a pair of
photons from a classical light beam of polarization $\vec M$ gets
projected onto the singlet state reads:

\bea
Prob(singlet)&=&Tr(\rho\otimes\rho\cdot P_{singlet})\\
&=&\frac{1-\vec M^2}{4}=\frac{1-DOP^2}{4}\label{DOPmeas}
\eea

The coherent quantum measurement "projection onto the singlet
state" provides thus a {\it direct access} to the DOP. In section
II we present a measurement setup, inspired by quantum optics
experiments (projection onto the singlet state is useful, among
others, for the fascinating demonstration of quantum teleportation
\cite{teleport}), but extended into the classical domain using
nonlinear optics. However, before this we would like to present an
example where a direct and fast DOP measurement is of great
practical value.

Polarization Mode Dispersion (PMD) is presently one of the main
limitations to high bit-rate fiber optics communication \cite{K}.
Consequently, the telecom industry aims at developing
compensators. This road has been taken successfully to fight
against chromatic dispersion. However, contrary to chromatic
dispersion, PMD is a statistical quantity which fluctuates on
various time scales, down to microseconds in the worst case.
Hence, any PMD compensator needs a fast feedback parameter.
Ideally, this parameter should be the Bit Error Rate (BER).
However, today's BER specifications of $10^{-9}$, or even
$10^{-12}$, impose much too long measurement times, even at bit
rates of tens of gigabits per second. An often proposed
alternative to the BER as feedback parameter is the DOP
\cite{PMDcomp}. Indeed, when PMD affects the transmission of light
pulses, then, in first order, one part of the pulse travels
slightly faster than the other, though they do still overlap.
Hence, the DOP during this overlap is the desired feedback
parameter. Clearly, in this case the depolarization is never due
to mixtures of spatial modes and the time fluctuations, e.g. from
one pulse to another, do not represent the physical quantity of
interest. This is a clear example where a direct and fast
measurement of the DOP is needed. In the frequency domain PMD can
be understood as follows. The light fields contains three dominant
optical frequencies, the carrier and the carrier $\pm$ the
modulation frequency. Each of these wavelengths undergo slightly
different polarization evolutions, hence the depolarization of
interest is clearly due to the third cause listed in the
introduction. For frequency modulations from giga- to terabits per
second, the wavelengths differences range from 8pm to 8nm.

\section{II Experimental setup}
The experimental implementation of the "projection onto the
singlet state" measurement is presented in Fig. 1. The idea is to
coherently combine two stages of parametric upconversion, using
$\chi^2$ type II nonlinear crystals. In the first stage, the
phase-matching is such that a photon from the shorter range of the
spectrum and one from the longer range are upconverted to a photon
in a horizontal polarization state. The second stage is rotated by
$90^{\circ}$, and consequently, the upconverted photon is
vertically polarized. The upconverted photons then pass a linear
polarizer at $45^{\circ}$, which erases the information where they
were created. Hence, the two processes add coherently. Depending
on the phase between the two stages, controlled by tilting two
birefringent plates, the overall intensity of the upconverted
signal corresponds to the desired "singlet-fraction", and is
consequently a measure for the DOP (Eq \ref{DOPmeas}).
%\begin{center}
\begin{figure}
     % Requires \usepackage{graphicx}
     \includegraphics[width=8 cm]{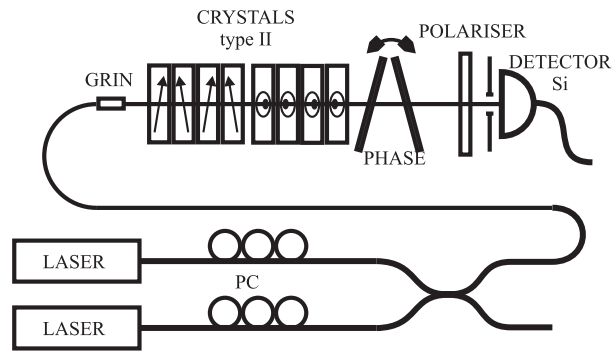}
     \caption{Diagram of the set-up. The two walk-off compensated
     stages of four nonlinear crystals are turned by
     90$^\circ$ with respect to each other. PC: polarization controller;
     GRIN: graded-index lens.}
     \label{graph1}
\end{figure}
%\end{center}
Note that the probability for upconvertion is
important during a time interval given by the coherence time of
the pump photons (position uncertainty). This means that the
signal amplitude at a given moment comes from pump fields averaged
over their coherence time. According to this "response time" of
the non-linear interaction, the outcome of our DOP-meter is the
"instantaneous" DOP as defined in the introduction.

A preliminary investigation using only two, orthogonally
orientated crystals \cite{DOPJMO2} showed that an undesired
phase-matching condition co-exists for photons with little
wavelength difference. For example, in the same crystal, the two
nonlinear interactions $((H_{1},V_{2})\rightarrow H_{3})$ and
$((V_{1},H_{2})\rightarrow H_{3})$ are possible. This poses a
serious limitation to the scheme. The wavelength separation under
which this detrimental phenomenon appears is determined by the
phasematching acceptance of the crystal. Hence, the narrower the
wavelength acceptance of the nonlinear crystals, the better,
contrary to the typical use of such crystals. To reduce the
wavelength acceptance, we can use longer crystals or choose
materials having better characteristics. Promising candidates as
GaSe, HgS (Cinnabar) , or Banana are however hard to fabricate or
difficult to manipulate. We therefore decided to stay with KTP,
but to increase the crystal length. This leads to a spatial
walk-off problem, limiting the effective length for SFG to well
below the physical crystal length. Usually this is dealt with by
adding linear birefringent crystals for compensation. Here, we
compensate the walk-off using a second nonlinear birefringent
crystal. As is described in \cite{Zondy}, two identical nonlinear
crystals are combined so that their walk-off angles are opposite
and the waves generated in both are in phase. To realize the
desired effective length, we use stages consisting of 4 KTP
crystals each, hence our set-up contained eventually 8 nonlinear
crystals in series. This is an interesting result in itself, since
recently many experiments presented configurations using just
pairs of nonlinear crystals \cite{photonpairs}.

A structure of four 3mm KTP elements gives an effective length of
almost 12mm, thereby reducing the wavelength acceptance by 4
compared to a 3mm crystal as used in \cite{DOPJMO2}. The expected
wavelength phasematching acceptance becomes 4.5nm, making it
possible to realize a projection onto the singlet state for
wavelengths separated by $\sim$1.5nm only. Notice that the spatial
walk-off is totally compensated for, so contrary to normal
crystals, the spatial modes of $\lambda_{1}$ and $\lambda_{2}$ are
as well overlapped before the second stage as before the first
one. This favorizes both identical conversion efficiencies in both
stages and a better spatial overlap of the created waves.

\section{III Results}
In this section we demonstrate the performance of our projection
on the singlet state with the 8 KTP crystals. To test the set-up,
we use a source composed of two lasers, one at the wavelength
$\lambda_{1}$ and the other at $\lambda_{2}$ (figure 1). Mimicking
PMD, the polarization of each wavelength is adjusted separately
with polarization controllers. The DOP of such a source is given
by $[(I_{1}+I_{2})^2-4I_{1}I_{2}sin^2\varphi]^{1/2}/(I_{1}+I_{2})$
where $2\varphi$ is the angle between the states of polarization
of $\overrightarrow{M}(\lambda_{1})$ and
$\overrightarrow{M}(\lambda_{2})$ (Poincare sphere). With this
source, it is very simple to study the response of our system for
many configurations. In the following, we concentrate on the case
$\lambda_{1}=1552nm$ and $\lambda_{2}=1554nm$. Similar results
were obtained for larger wavelength separations.

First, we characterize the quality of our projection onto the
singlet state. For any input polarization combination, the output
of our device has to be proportional to $1-DOP^{2}$ (Eq. 3). To
well cover the possible inputs with a reasonable number of
measurements, we choose polarization states on three orthogonal
great circles of the Poincare sphere. For each great circle,
$\overrightarrow{M}(\lambda_{1})$ is set to 5 polarization states
separated by $40^{\circ}$. For each of those states,
$\overrightarrow{M}(\lambda_{2})$ is chosen on the same circle so
that $2\varphi=0, 10,..., 90^{\circ}$, corresponding to ten
different values for the DOP. The measured data are shown in
figure 2, where the values obtained from the different circles are
represented by different symbols (squares, circles, and
triangles). Due to the choice of polarization states, for each
circle we have 5 points for a given DOP (corresponding to the 5
different absolute input polarization directions). As expected,
the detected intensity reflects the DOP of our source, and is
quasi-independent of the absolute polarization states of
$\lambda_{1}$ and $\lambda_{2}$. The residual fluctuations
observed for a given DOP value are due to misalignments of the
set-up. Specifically, the small variations for a DOP of 1 are
essentially due to a slightly reduced visibility of the
interferences between the two waves from the two stages (see
\cite{DOPJMO2} for more details). We can estimate a visibility of
more than 96$\%$. This is achieved thanks to a proper spatial
overlap of the modes created in the two stages due to walk-off
compensation in the crystals. If we estimate the precision of our
measurement with the standard deviation of the fluctuations, the
error of our device on the determination of the DOP is a few
percent for a DOP close to 1 and about 15$\%$ for a totally
depolarized source. Figure 2 also shows the mean values for a
given DOP (open circles). They follow very well the linear-law
predicted by the theory (solid line).
%\begin{center}
\begin{figure}
     % Requires \usepackage{graphicx}
     \includegraphics[width=8 cm]{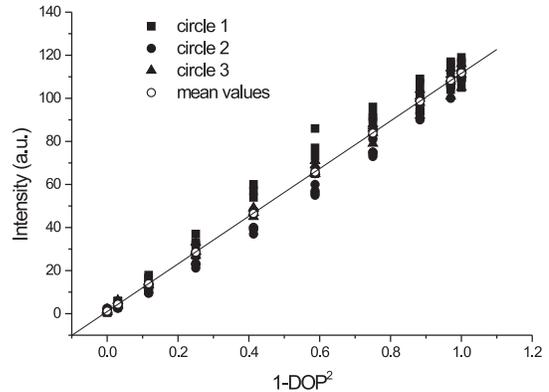}
     \caption{Measured intensity of the projection onto the
     singlet state as a function of $1-DOP^{2}$ for
     $\lambda_{1}=1552nm$ and $\lambda_{2}=1554nm$.}
     \label{graph2}
\end{figure}
%\end{center}

So far the analyzed signal was constant in time. In order to
demonstrate that we really measure the "instantaneous" DOP, a
source with constant DOP but rapidly fluctuating state of
polarization is required. We realize this by shaking the fiber
linking the source to the DOP meter (fiber after the coupler in
figure 1). This leads to variations in the birefringence axis
direction and Berry's phase in this fiber, and consequently the
polarization states $\overrightarrow{M}(\lambda_{1})$ and
$\overrightarrow{M}(\lambda_{2})$ will strongly fluctuate in time.
If the amount of birefringence is small enough compared to the
wavelength difference $\lambda_{1}-\lambda_{2}$, the relative
polarization angle $\varphi$ between
$\overrightarrow{M}(\lambda_{1})$ and
$\overrightarrow{M}(\lambda_{2})$ (i.e. the DOP) is conserved even
when agitating the fiber. In our experiment, we are manually
moving the fiber leading to a time scale of the polarization
fluctuations of $\sim$100ms. Accordingly, an integration time of a
few seconds is chosen in order to be sure that the polarization
state strongly fluctuates during this time interval. Figure 3
shows corresponding results for 3 different values of the DOP
(open symbols, integration time 10s). The fiber was not shaken for
the first and last measurement points to have 2 reference values.
As can be seen, the same values for the DOP are obtained when
shaking the fiber. This clearly demonstrates the projection onto
the singlet state does indeed give the "instantaneous" DOP.

To illustrate that this is not the case for the standard
measurement techniques, we repeated the measurement using a
polarimeter with 10s integration time (PAT-9000, Profile). On the
first and last point, we measure the same value as with the
singlet state projection. But when the fiber is shaken the
measured value of the DOP strongly decreases and also fluctuates
somewhat. This behavior is observed both for 10s (figure 3) and 1s
integration times. Clearly, the DOP is no longer measured
correctly. Note that although a polarimeter can integrate much
faster than 1 second (e.g. 33ms for the PAT-9000), the same
problem will be observed for fluctuations of the order of
milliseconds as they can occur for PMD.
%[Insert figure 3 about here]
%\begin{center}
\begin{figure}
     % Requires \usepackage{graphicx}
     \includegraphics[width=8 cm]{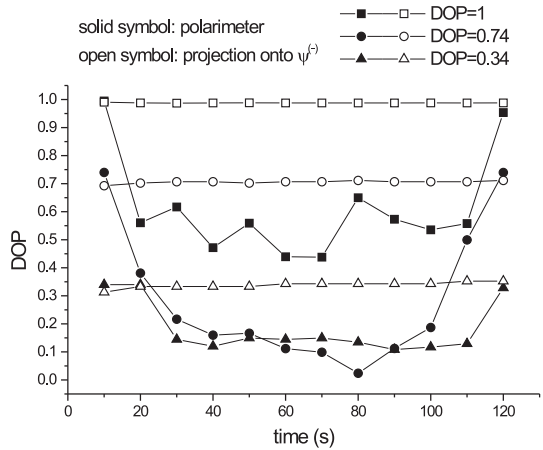}
     \caption{DOP measured with our device (open symbols) and
     with a polarimeter (solid symbols) as a function of time.
     The DOP of the source is constant but its polarization
     state fluctuates with time, except for the first and
     last measurement points of each curve where it was fixed.}
     \label{graph3}
\end{figure}
%\end{center}
\section{conclusions}

A concrete application of a coherent quantum measurement has been
realized: a DOP-meter. It is based on the projection onto the
singlet state, and allows to measure the instantaneous DOP in a
direct way. This is different from the standard, indirect method
of DOP evaluation (polarimetric technique) where the DOP is
averaged over the integration time of the detection, which is
typically longer than the coherence time of the signal to be
measured. Consequently, for a signal with temporally fluctuating
polarization only the first method gives the correct DOP.

Experimentally the projection onto the singlet state is realized
exploiting up-convertion in two type II nonlinear crystals. In
order to increase the efficiency of the process and to be able to
measure signals with narrow spectra, the effective crystal length
should be large. We achieved this by stacking 2x4 KTP crystals of
3mm length in a walk-off compensation arrangement, giving an
effective length of almost 12mm for each of the two stages. With
this compensation technique, we obtained a high quality DOP
measurement for wavelengths separated by 2nm. Further, we
demonstrated that the projection onto the singlet state gives
indeed the "instantaneous" DOP. For a signal with temporally
fluctuating polarization we still obtained the correct value,
whereas this was not the case for a standard polarimetric
measurement.

%\section{Acknowledgements}
\textit{\textbf{Acknowledgements:} Financial support from the
Swiss OFES in the frame of the COST 265 project, EXFO Inc (Vanier,
Canada), and the Swiss NCCR "Quantum photonics" are acknowledged.}

\bigskip
\small{$^*$Matthieu.legre@physics.unige.ch}

\end{document}